\documentclass[prl,aps,twocolumn,epsfig]{revtex4}
\usepackage{graphicx}
\usepackage{dcolumn}
\usepackage{bm}
\usepackage{hyperref}
\bibpunct{(}{)}{,}{s}{,}{,}
\makeatletter

\newcommand{\Rmnum}[1]{\expandafter\@slowromancap\romannumeral #1@}
\makeatother
\begin{document}
\title {The Nature of Electronic States in Atomically Thin MoS$_2$ Field-Effect Transistors}
\author{Subhamoy Ghatak,$^1$}
\email{ghatak@physics.iisc.ernet.in}
\author{Atindra Nath Pal,$^1$ and Arindam Ghosh$^1$}
\vspace{1.5cm}
\address{$^1$ Department of Physics, Indian Institute of Science, Bangalore 560 012, India}

\begin{abstract}
We present low temperature electrical transport experiments in five field effect transistor devices consisting of monolayer, bilayer and
trilayer MoS$_2$ films, mechanically exfoliated onto Si/SiO$_2$ substrate. Our experiments reveal that the electronic states in all films are
localized well up to the room temperature over the experimentally accessible range of gate voltage. This manifests in two dimensional (2D)
variable range hopping (VRH) at high temperatures, while below $\sim 30$~K the conductivity displays oscillatory structures in gate voltage
arising from resonant tunneling at the localized sites. From the correlation energy ($T_0$) of VRH and gate voltage dependence of conductivity,
we suggest that Coulomb potential from trapped charges in the substrate are the dominant source of disorder in MoS$_2$ field effect devices,
which leads to carrier localization as well.
\vspace{10 mm}

{\bf KEYWORDS:} Dichalcogenides. Field effect transistor. MoS$_2$. Localization. Mott variable range hopping. Resonant tunneling. Charge
impurity scattering.
\end{abstract}


\maketitle

MoS$_2$-based dichalcogenides have recently been of renewed interest due to the possibility of creating atomically thin semiconductor membranes
for a variety of applications.~\cite{photolumi,tonyprl,Ramanacsnano,single} Being a layered compound, with a weak van der Waal interaction
between the layers, MoS$_2$ can be exfoliated like graphene on insulating substrates. This has recently led to the fabrication of single-layer
MoS$_2$ field effect transistor that has a very high on-off ratio due to a finite bandgap.~\cite{single} It has been demonstrated that the
bandgap is indirect ($\approx 1.2$~eV) for multilayer MoS$_2$ films but direct ($\approx 1.8$~eV) for single atomic layer,~\cite{Ramanacsnano}
which may lead not only to low-power dissipation electronic devices, but also new possibilities in energy harvesting designs. The existence of
bandgap can also have serious implications on the charge transport and nature of disorder in MoS$_2$ films, affecting its ability to screen
external potential fluctuations.~\cite{atindaprl,ouracsnano} In fact, the mobility ($\lesssim 200$~cm$^2$/V.s) of charge carriers in single
layer MoS$_2$ devices is much lower than graphene , which has been attributed to the absence of a bandgap in pristine 2D layers of graphene. Our
objective here is to explore the nature of disorder, and hence that of the electronic states, from the low temperature electrical transport in
MoS$_2$ films, when the film thickness is downsized from few to a single molecular layer.

Often, disorder in low dimensional electron systems arises from extraneous sources, such as local charge distribution that induces a random
Coulomb potential on the electrons. This, for example, can be the remote dopant ions in modulation-doped III-V semiconductor,~\cite{efrosprb} or
charges trapped in the substrate in case of graphene.~\cite{adamremote} At low carrier densities the screening of the random Coulomb potential
becomes weak, causing carriers to localization and/or charge distribution to become inhomogeneous.~\cite{ehpuddle,Matthiasprbrapid} This would
not only have a direct impact on the transport, but also on the response of the system to various stimuli including light, stress $etc$.
Although a similar charge trap-mediated transport has been suggested in thick nanoscale patches of MoS$_2$,~\cite{nanopatch} the microscopic
picture is far from clear. It is also not known if such a picture would be valid in mono or very few layer MoS$_2$ devices. In this context, our
experiments with MoS$_2$ devices of different thicknesses (see Table I) reveal that electrons are strongly localized in all cases which manifest
in variable range hopping transport and an inhomogeniety in charge distribution that results in local transport resonances at low temperatures.
We suggest that localization is probably due to strong potential fluctuations induced by the randomly occupied charge traps that are located
primarily at the MoS$_2$-substrate interface.

\begin{table*}
\caption{\label{table1}Details of the devices:}
\begin{ruledtabular}
\begin{tabular}{cccccc}
 Device & Number of layer & Contact material & $V_{ON}$\footnote{in V} & Device area (L$\times$W)\footnote{both dimensions in $\mu$m} &  Mobility\footnote{in cm$^2$/V-s near room temperature}\\ \hline
 MoS1La & 1 & Ti/Au & 32 & $4 \times 3$ & 1  \\
 MoS1Lb & 1 & Au & 15 & $2 \times 2.5$ & 5  \\
 MoS1Lc & 1 & Au & -5 & $5 \times 8$ & 12 \\
 MoS2L & 2 & Au & -2 & $2.8 \times 2.5$ & 20 \\
 MoS3L & 3 & Ti/Au & -25 & $4 \times 16$ & 10 \\

\end{tabular}
\end{ruledtabular}
\end{table*}

\setcounter{secnumdepth}{-1}
\section{RESULTS AND DISCUSSION}

Devices were prepared by standard mechanical exfoliation of bulk MoS$_2$ on 300 nm SiO${_2}$ on $n^{++}$ doped silicon substrate using scotch
tape technique.~\cite{novopnas,novo} The flakes were identified using optical microscope and characterized $via$ Raman spectroscopy and atomic
force microscopy (AFM). We present detailed experiments on five devices (See Table I) of different film thicknesses. In Fig.~1c we show Raman
spectra for bulk, tri and single layer MoS$_2$ films. We focus on the E$^1_{2g}$ and A$_{1g}$ modes which have been shown to be sensitive to the
number of atomic layers.~\cite{Ramanacsnano} The position of both modes agree well with recent investigation of Raman spectroscopy in thin
exfoliated MoS$_2$ films. The separation of E$^1_{2g}$ and A$_{1g}$ peaks were found to be 23~cm$^{-1}$, 21~cm$^{-1}$ and 16 -18~cm$^{-1}$ for
tri, bi and single layer respectively.  For further confirmation of Raman data, we determined the thickness of flakes using contact mode AFM. A
line scan across the edge of a single layer flake (Fig.~1d) shows a step of $\approx 0.7$~nm which compares very well with the thickness of the
single MoS$_2$ layer ($\approx 0.65$~nm).

The electrical contacts, designed with electron beam lithography, consisted of thermally evaporated Ti/Au or Au films. The optical image of the
MoS3L device is shown in Fig.~1a. All measurements were carried out in cryostats under high vacuum ($10^{-6}$~mbar) condition. In all devices
the gate voltage ($V_{BG}$) was applied only at the doped silicon backgate (see Fig.~1b). Measurements were primarily two-probe current
measurement using lockin technique due to very high resistance of these systems, although four-probe devices were fabricated as well. We found,
at high doping concentration, the contact resistance was negligible near room temperature, but increases to about half of sample resistance
below 100K. Detailed $I_{DS}-V_{DS}$ measurements, where $I_{DS}$ and $V_{DS}$ are the drain-source current and bias respectively, were
conducted to characterize the electrical contacts (see supplementary information). At low voltages ($|V_{DS}|$ $\lesssim$ $300$~mV),
$I_{DS}-V_{DS}$ at all $V_{BG}$ and near room temperature, were linear for both Ti/Au and Au deposited samples although we have got better
linear contact with only Au. These results bear close resemblance to the characteristics reported recently for high mobility MoS$_2$
devices.~\cite{single} As shown in the supplementary information, $I_{DS}-V_{DS}$ characteristics become nonlinear at large $V_{DS}$,
particularly at low temperatures ($T$), although we attribute this to the insulating nature of the devices which causes the nonlinearity. The
symmetric nature of $I_{DS}$-$V_{DS}$ around $V_{DS} = 0$ enables us to eliminate any possibility of Schottky contact in our operating $V_{DS}$
range. This is supported by the observed magnitude of the differential carrier mobility $\mu (= (1/C)\times d\sigma/dV_{BG})$, where $C$ is the
gate capacitance per unit area (here $1.2 \times 10^{-4}$ F/m$^2$ for 300 nm SiO$_2$), and $\sigma(= (L/W)\times I/V_{DS})$ is the linear
conductivity at low $V_{DS}$. $L$ and $W$ are the length and width of the MoS$_2$ channel. In both two and four-probe geometry, we obtained
similar values of mobility, which are typical values reported for MoS$_2$ transistors~\cite{single} on SiO$_2$ substrate (See Table I).

In Fig.~2 we show the variation of $\sigma$ in MoS3L and MoS1La as a function of $V_{BG}$ over a range $\sim 10$~K to 280~K. The conduction was
achieved predominantly in the positive $V_{BG}$ regime implying that the MoS$_2$ films were intrinsically $n$-type.~\cite{nanopatch} The doping
was higher in MoS3L which required a negative $V_{BG}$ to pinch-off completely. Below $250$~K both devices were strongly insulating at all
$V_g$. A weak metal-like behavior was observed at $T> 250$~K and very high doping (large positive $V_{BG}$) for most of the samples. Such a
behavior, which was stronger in MoS3L, was found to be connected to a decrease in $\mu$ with increasing $T$ in this regime (inset of Figs.~2a
and 2b). This can be attributed to enhanced scattering of electrons by phonons at high temperatures.~\cite{syntheticmetal} At low $T$, both
$\sigma$ and $\mu$ drop rapidly with decreasing $T$. Such an insulating behavior was observed in multilayer nanopatches of MoS$_2$ (thickness: 8
$-$ 35~nm) as well,~\cite{nanopatch} with an apparent activated behavior of $\sigma$ over a rather limited range of $T$. This was explained by
invoking a dense distribution of trap states, which acted as an ``impurity band'', although the origin or the physical location of such traps
are unclear.

In contrast to the nanopatches,~\cite{nanopatch} the $T$-dependence of $\sigma$ in our mono and trilayer MoS$_2$ devices are not activated, but
there are two distinct regimes in the $T$-variation in $\sigma$ (Figs.~3a and 3b): The high-$T$ regime ($T \gtrsim 30$~K), where $\sigma$
increases rapidly with increasing $T$, and second, the low-$T$ regime ($T \lesssim 30$~K), where the variation in $\sigma$ weakens considerably
at most $V_{BG}$ in both devices (except for MoS1La at low $V_{BG}$, where the weakening of $\sigma$ sets in at higher $T$ (Fig.~3b)). We find
that in the high $T$ regime, the variation of $\sigma$ with $T$ can be modeled very well in terms of variable range hopping (VRH) transport
rather than the thermally activated behaviour with
\begin{equation}
\label{eq1}\sigma =\sigma_0(T)\exp[-(T_0/T)^{\frac{1}{d+1}}]
\end{equation}
where $T_0$ and $d$ are correlation energy scale and dimensionality,~\cite{mottnoncrystalline,sigmatpowerm} respectively, and $\sigma_0 = AT^m$
with $m \approx 0.8 - 1$. The agreement of the data to VRH transport with $d = 2$ indicates the electron transport in atomically thin MoS$_2$ to
occur in a wide ($\gg k_BT$) band of localized states, rather than direct excitation to conduction band minimum or mobility edge from the Fermi
energy as suggested for the nanopatches.~\cite{comment} The VRH transport in $\sigma$ also results in $\ln\mu \propto T^{-1/3}$ in two
dimension.~\cite{syntheticmetal} This is confirmed in the inset of Fig.~3b for the MoS1La device. The magnitude of $T_0$ decreases rapidly as
$V_{BG}$, or equivalently, the Fermi energy $E_F$, is increased. Such a behavior is common to strongly localized 2D electron
systems,~\cite{pollitt,arindamhopping} and implies that $E_F$ is located in the conduction band tail.

To understand the weakening of $\sigma$ at $T \lesssim 30$~K, we have magnified this regime for both MoS3L (Fig.~3c) and MoS1La (Fig.~3d). In
both cases, the variation of $\sigma$ with $V_{BG}$ becomes nonmonotonic, and displays several peaks which become progressively well defined as
$T$ is reduced. The peaks are highly reproducible, and stable even at $T \sim 30 - 40$~K, indicating that random fluctuations due to
interference of hopping paths are unlikely to cause them. Resonant tunneling at the localized states in disordered mesoscopic semiconductors is
known to result in strong reproducible peaks in $\sigma$ at low temperatures.~\cite{fowler,arindampeak} In the presence of multiple overlapping
resonances, $T$ dependence of $\sigma$ weakens as observed in our data.~\cite{arindampeak} However, confirmation of this scenario can be
obtained by shifting the resonance peaks using finite $V_{DS}$. For this, we focused on a small interval of $V_{BG}$ ($8 - 35$~V) near pinch-off
where a number of isolated resonances could be identified. In the $(V_{BG}, V_{DS})$ plane, this leads to diamond-like pattern in differential
conductivity $dI/dV_{DS}$ (Inset: Fig 3c). The occurrence of transport resonances indicates a rather inhomogenous charge distribution in MoS$_2$
films, possibly puddles of charge near conduction threshold, through which charging events at the localized states couple to the metal contacts.

We now turn into the key issue here that concerns the origin of localized states in ultrathin MoS2 films. This requires an understanding of the
origin of disorder in such systems, for which we first examine the values of $T_0$. However, to compare $T_0$ for different devices, we define a
device-specific reference voltage $V_{ON}$ close to the ``pinch-off'' voltage in $\sigma ~ vs ~ V_{BG}$ curve, so that the difference $\Delta
V_{BG} = V_{BG} - V_{ON}$ is proportional to $E_F$ or number density $n$. In Fig.~4a, we have plotted the variation of $T_0$ as a function of
$\Delta V_{BG}$ for all the devices. The striking feature here is the close agreement of $T_0$ in both absolute magnitude and energy over nearly
three decades, irrespective of independent preparation of devices, varying layer number, mobility and device geometry $etc$. This indicates a
very similar disorder landscape in all devices that reflects comparable magnitude and energy dependence of localization length $(\xi)$ and
density of states $D(E)$. Disorder arising from defects in bulk of the MoS$_2$ films are unlikely to explain the insensitivity of $T_0$ to
number of layers since screening of impurities and density of defect in bulk are expected to strongly influence the density of localized states.
In stead, our data indicates a common external origin of disorder, such as the trapped charges in the substrate. This is also supported by
recent transport experiments,~\cite{single} where higher mobility of thin MoS$_2$ flakes could be achieved by changing electrostatic environment
alone. Indeed a charge trap-induced disorder can readly explains the observed magnitude of $T_0$. To illustrate this we take $\xi$ as the
typical size of the puddles, which for MoS3L can be roughly estimated to be $\xi \sim 8$~nm from the charging energy ($\sim 90$~meV) at $V_{BG}
\approx 23$~V (corresponding to $\Delta V_{BG} \approx 48$~V) (See inset of Fig.~3c). Taking $D(E) \sim 4\times10^{12}$~eV$^{-1}$cm$^{-2}$ as
the typical surface density of charge traps at SiO$_2$ interface,~\cite{nanopatch,jayaraman} and using $T_0 = \frac{13.8}{k_B\xi^2(E) D(E)}$, we
find $T_0 \simeq 6.2\times10^4$~K, which is in good agreement to the observed magnitude from VRH data (Fig.~4a).

This leads us to suggest that the physical origin of the localized states in ultra-thin MoS$_2$ films is connected to the random potential
fluctuations from the trapped charges at the MoS$_2$-SiO$_2$ interface. (See the schematic of Fig.~4b.) The screening of these trapped charges
will be poor due to the large bandgap of MoS$_2$ (unlike graphene) and hence can lead to a considerably long band tail. It is likely that the
interfacial traps are randomly occupied during processing of the devices, predominantly $via$ transfer of electrons from the exfoliated prestine
MoS$_2$ layers, and subsequently form the frozen disorder landscape since most experiments are conducted at low $T$.

Finally to confirm the charge impurity induced disorder, we have examined the nature of scattering of carriers by defects at high $V_{BG}$ and
$T$ so that the electron wavefunctions are nearly extended. If the main source of disorder arises from the randomly occupied interfacial traps,
one would expect the scattering to be dominated by charge impurity scattering, which for  two-dimensional electron systems with parabolic energy
bands will lead to,~\cite{sigmansqdassarma}

\begin{eqnarray}
\label{eq2} \sigma & \propto & n^2, \quad \mbox{bare Coulomb impurity}\\
& \propto & n, \quad \mbox{screened Coulomb impurity}
\end{eqnarray}

\noindent In Fig.~5a, b and c, we have shown the dependence of $\sigma$ on $\Delta V_{BG}$ ($\propto n$) near room temperature, for the single,
bi and tri layer MoS$_2$ devices, respectively. In all monolayer MoS$_2$ devices, as well as the bilayer (MoS2L) case, we find  $\sigma \propto
\Delta V_{BG}^2$ indicating scattering from nearly unscreened charged impurities. In the trilayer device (MoS3L), the variation in $\sigma \sim
\Delta V_{BG}^{1.6}$ is somewhat slower, indicating partial screening of the charge impurities. Assuming the electronic density of states to be
approximately one-tenth of the free electron density of states at maximal doping $(\sim 5\times10^{12}/cm^2)$ used in our experiment the Debye
screening length in our devices can be estimated to be $\sim 1.5-2 ~nm$, which is nearly three molecular layers of MoS$_2$. This readily
explains the bare charge impurity scattering in single and bi-layer MoS$_2$, while charge impurities are partially screened for the trilayer
device.

It is then natural to draw an analogy of our findings to other heavily researched exfoliated atomic scale transistors in particular, graphene
and topological insulators. The ubiquity of surface trap states probably constitutes a generic source of disorder in such ultra-thin field
effect devices. Reducing substrate traps, for example, by using crystalline substrates such as graphene on boron nitride, may improve the
quality of these systems considerably. A suspended device, as in case of graphene, could also lead to extremely high mobilities.

\section{CONCLUSION}

We have studied low temperature electrical transport in mono, bi and trilayer MoS$_2$ transistor exfoliated onto Si/SiO$_2$ substrates. We find
that the electrons in all cases are localized well upto room temperature at most gate voltages, and displays variable range hopping transport as
temperature is lowered. We showed that the disorder is likely to arise from Coulomb potential of randomly distributed charges at the
MoS$_2$-SiO$_2$ interface, and hence highly improved devices should be possible with appropriate substrate engineering.

\section{METHODS}
\noindent{\bf Device Fabrication} \vspace{0.3cm}

MoS${_2}$ flakes were exfoliated from bulk MoS$_2$ (SPI Supplies) using scotch tape on SiO$_2$ (300~nm)/n$^{++}$ Si wafer. To keep the disorder
level comparable, the wafers were thoroughly cleaned by standard RCA cleaning followed by acetone and isopropyl alcohol cleaning in ultrasonic
bath. The flakes with typical linear dimensions ranging from 2 $\mu$m to 20 $\mu$m were identified by Olympus BX51 optical microscope. Raman
spectrum are recorded using WITEC confocal (X100 objective) spectrometer with 600 lines/mm grating, 514.5 nm excitation at a very low laser
power level (less than 1 mW) to avoid any heating effect. The AFM measurements were carried out in contact mode with a NT-MDT NTEGRA AFM
instrument. Ti(10 nm)/Au(40 nm) or Au(40 nm) contacts were defined using standard electron beam lithography followed by thermal evaporation and
lift off in hot acetone. No Ar/H$_2$ annealing was done in any of our devices.

\emph{Acknowledgement.} We acknowledge Department of science and Technology (DST) for a funded project. S.G. and A.N.P. thank CSIR for financial
support.

\emph{Supporting Information Available.}~Detail room temperature and low temperature drain source characteristics are presented for trilayer and
monolayer devices in Figure S1. The temperature dependent data for Mott type variable range hopping for MoS1Lc is shown in Figure S2 along with
the calculation of VRH slope. This material is available free of charge $vi$a the Internet at \url{http://pubs.acs.org}.

\newpage

\begin{figure*}[h]
\includegraphics[width=1\linewidth]{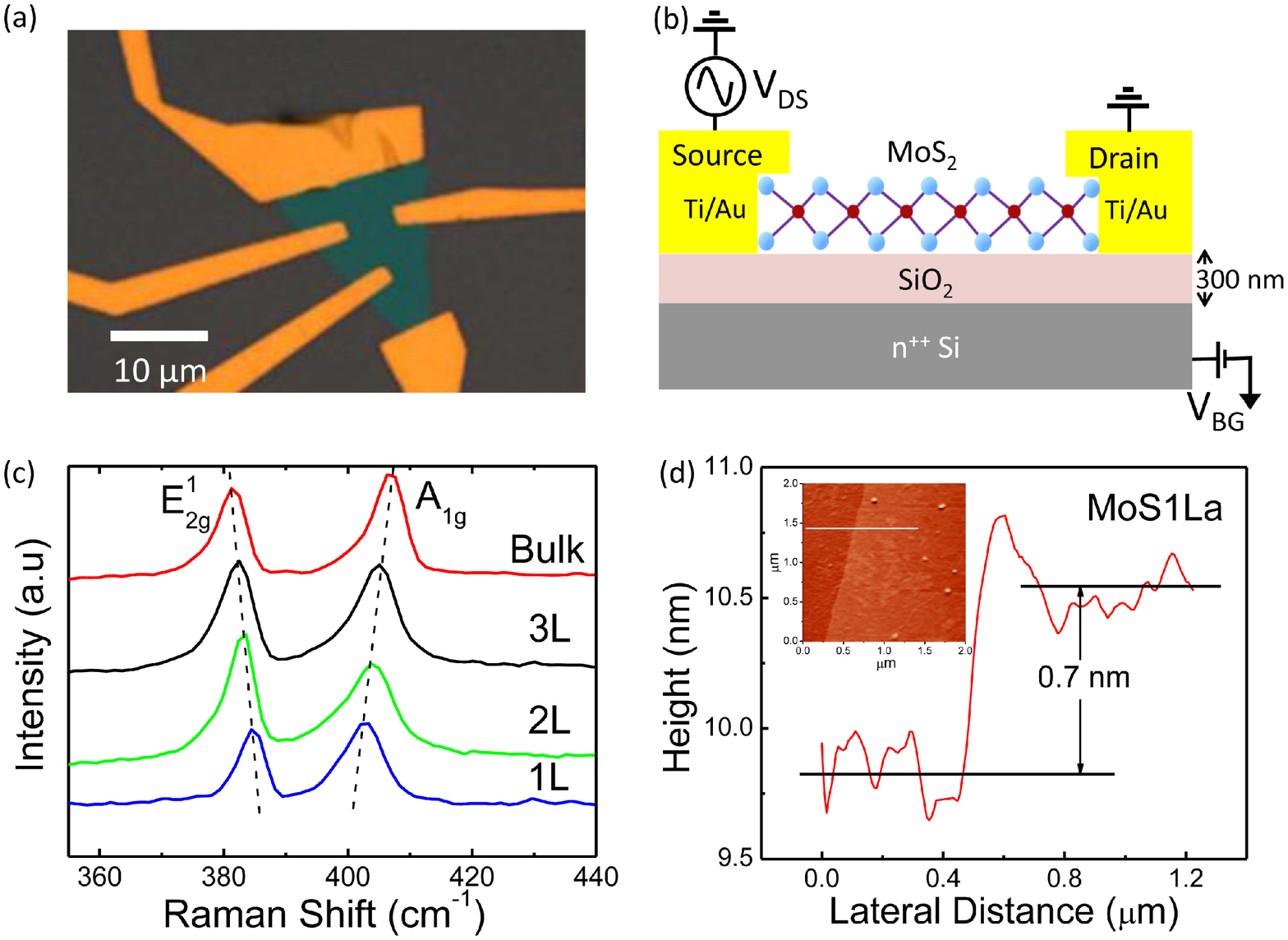}
\caption{(a) Optical micrograph of a typical MoS$_2$ device. (b) Schematic of a single layer MoS$_2$ field effect transistor. (c) Raman spectrum
of the bulk, tri, bi and single layer MoS$_2$ films on Si/SiO$_2$ substrate. (d) Thickness scan along the white line across the boundary of the
single layer MoS$_2$ in the inset. Inset: High resolution atomic force microscopy (AFM) image of single layer MoS$_2$ film on SiO$_2$
substrate.}
\end{figure*}

\begin{figure*}[h]
\includegraphics[width=1\linewidth]{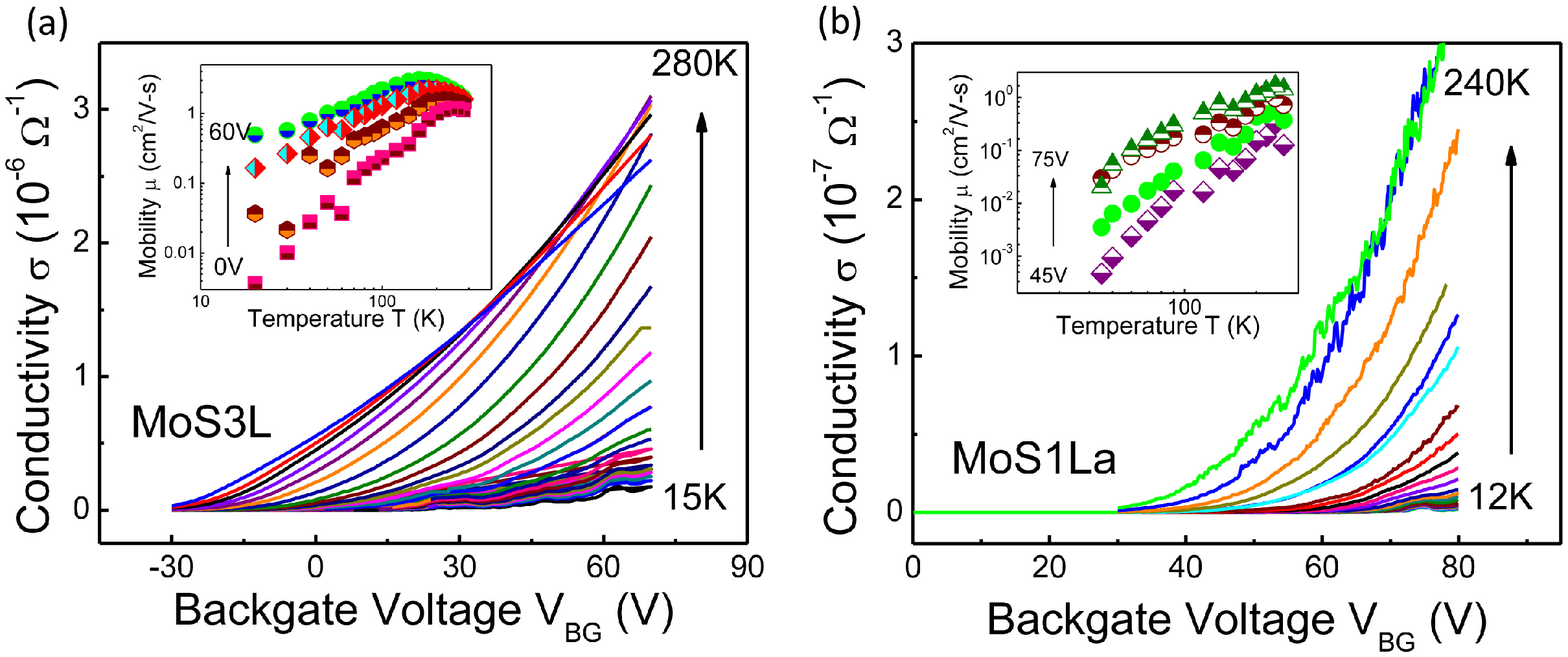}
\caption{Conductivity $\sigma$ as a function of back gate voltage ($V_{BG}$) at various temperatures for (a) MoS3L with $V_{DS}$ = 4~mV and (b)
MoS1La with $V_{DS}$ = 100~mV. Insets show corresponding field effect mobility $\mu$ vs temperature $T$ at
 different gate voltages $V_{BG}$, extracted from the linear fit of a small region around a particular $V_{BG}$ in the $\sigma$ vs $V_{BG}$ graph. }
\end{figure*}

\begin{figure*}[h]
\includegraphics[width=1\linewidth]{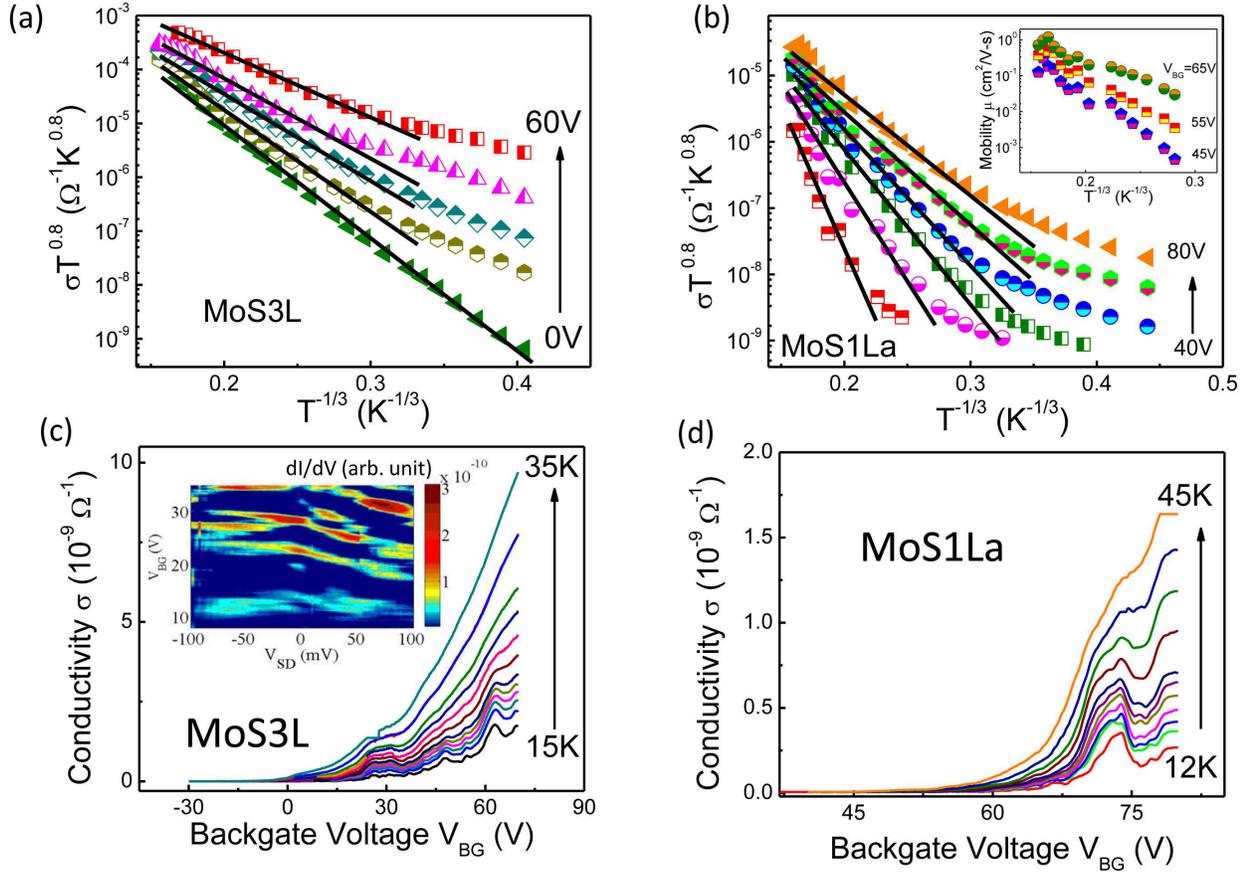}
\caption{Temperature dependence of conductivity ($\sigma$) and variable range hopping (VRH) at different backgate voltages, for (a) MoS3L
($V_{DS}$ = 4~mV) and (b) MoS1La ($V_{DS}$ = 100~mV). The solid black lines are the linear fit to the data indicating VRH behavior in 2D MoS$_2$
film. Inset in Fig.~3b shows variation of mobility $(\mu)$ with $T^{-1/3}$ for the single layer device. (c)-(d) Reproducible conductance
oscillations with back gate voltage ($V_{BG}$) at low temperature are shown for MoS3L and Mos1La respectively. Inset of Fig.~3c shows 2D map of
the differential conductance $dI/dV_{DS}$ of MoS3L as a function of back gate voltage ($V_{BG}$) and source-drain bias voltage ($V_{DS}$)
obtained at 12K. }
\end{figure*}

\begin{figure*}[h]
\includegraphics[width=1\linewidth]{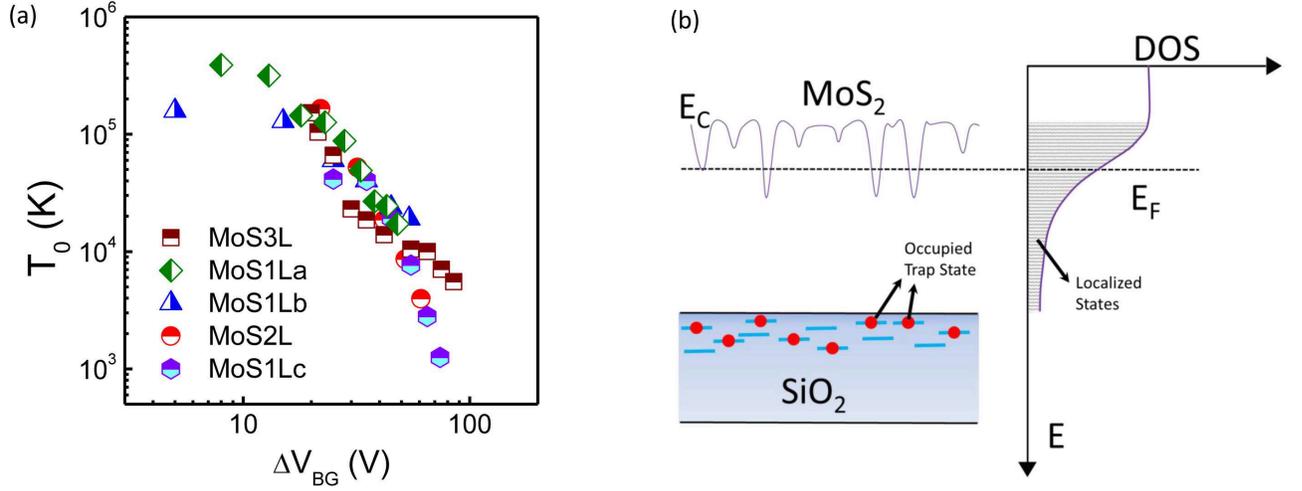}
\caption{(a) $T_{0}$, extracted from VRH slope for five different devices are plotted as function of $\Delta$$V_{BG}$ (= V$_{BG}$-V$_{ON}$). (b)
Schematic representation of the fluctuations in conduction band of MoS$_2$ thin films, arising due to proximity of the trapped charges at
SiO$_2$/MoS$_2$ interface (left) leading to the band tail and localized states (right).}
\end{figure*}

\begin{figure*}[h]
\includegraphics[width=1\linewidth]{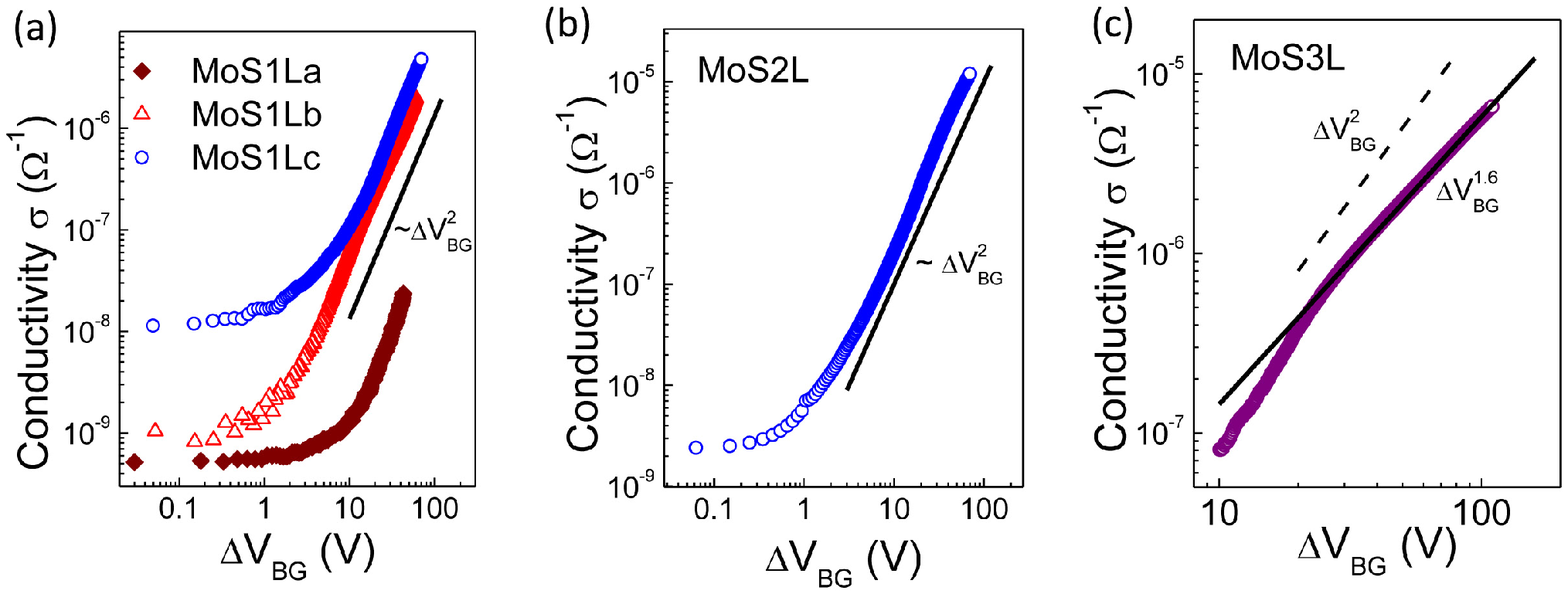}
\caption{Variation of conductivity $\sigma$ with $\Delta$$V_{BG}$ for (a) Single layer at 240K (diamond) and 300K (triangle, circle) (b) Bilayer
at 300K and (c) Trilayer at 280K devices.}
\end{figure*}


\begin{thebibliography}{30}
\bibitem{comment} The Efros-Shklovskii type variable range hopping with ln$\sigma\propto
  T^{-1/2}$~was also considered. However the data was found to have much better
  agreement with ln$\sigma\propto T^{-1/3}$ in almost all cases.
\end{thebibliography}
\end{document}